\documentclass[%
 reprint,
 amsmath,amssymb,
 aps,
]{revtex4-2}
\usepackage{mathptmx}
\usepackage{etoolbox}
\usepackage{xcolor}
\usepackage{float}
\usepackage{hyperref}
\usepackage{graphicx}
\usepackage{dcolumn}
\usepackage{bm}

\begin{document}

\preprint{APS/123-QED}

\title{Strong Electron-Phonon Coupling and Multiband Superconductivity in Hexagonal BP$_3$ Monolayer}

\author{Jakkapat Seeyangnok$^{1}$}
 \email{jakkapatjtp@gmail.com} 
\author{Udomsilp Pinsook$^{1}$}%
 \email{Udomsilp.P@Chula.ac.th}
\affiliation{Department of Physics, Faculty of Science, Chulalongkorn University, Bangkok, Thailand.}%



\date{\today}

\begin{abstract}
We investigate the structural, electronic, and superconducting properties of a hexagonal BP$_3$ monolayer using first-principles calculations combined with anisotropic Migdal--Eliashberg theory. The optimized structure exhibits a stable, slightly buckled configuration, as confirmed by phonon dispersion analysis and \textit{ab initio} molecular dynamics simulations. The phonon spectrum indicates high-frequency vibrational modes associated with B--P bonding. Electronic band structure calculations reveal a multiband metallic state, with states near the Fermi level predominantly derived from $p_z$ orbitals of both boron and phosphorus atoms, forming two distinct Fermi surface sheets. The electron--phonon coupling is relatively strong, with a total coupling constant of $\lambda = 1.59$, dominated by low- and intermediate-frequency phonon modes. Solving the anisotropic Migdal--Eliashberg equations yields a superconducting transition temperature of $T_c = 9.7$~K. The superconducting state is characterized by a nodeless but anisotropic gap structure, exhibiting two distinct gap values of approximately 2.25 and 1.74~meV associated with different Fermi surface sheets. These results identify the BP$_3$ monolayer as a strongly coupled, multiband two-dimensional superconductor and provide insight into the role of orbital hybridization in electron--phonon-mediated superconductivity in low-dimensional systems.
\end{abstract}

\keywords{Two-dimensional materials, BP$_3$ monolayer, electron--phonon coupling, superconductivity, two-gap superconductivity, Migdal--Eliashberg theory, density functional theory, multiband systems}
\maketitle

\section{Introduction}
    Two-dimensional (2D) materials have attracted tremendous interest due to their rich physical properties and wide-ranging potential applications in nanoelectronics, energy devices, and quantum technologies~\cite{novoselov2004electric,geim2013van,novoselov20162d,xu2013graphene,manzeli20172d}. Since the discovery of graphene~\cite{novoselov2004electric}, a broad variety of 2D systems have been extensively explored, including transition metal dichalcogenides~\cite{wang2012electronics,manzeli20172d}, group-IV monolayers such as silicene and germanene~\cite{vogt2012silicene,li2014buckled}, as well as elemental and compound-based layered materials~\cite{xu2013graphene,novoselov20162d}. Among these, low-dimensional superconductors have emerged as a particularly important class, as reduced dimensionality can significantly enhance electronic correlations and electron--phonon interactions, potentially leading to unconventional superconducting behavior and novel quantum phases~\cite{uchihashi2017two,saito2016superconductivity,xi2016ising}.

    Phonon-mediated superconductivity in two-dimensional (2D) materials has been extensively investigated within the framework of Migdal--Eliashberg theory, where the intricate interplay between electronic structure and lattice vibrations governs the superconducting transition temperature ($T_c$)~\cite{migdal1958interaction,eliashberg1960interactions,giustino2017electron}. A variety of 2D systems, including doped graphene~\cite{profeta2012phonon}, borophene~\cite{gao2017prediction,penev2012polymorphism}, and transition metal dichalcogenides~\cite{ge2015superconductivity,ugeda2016characterization}, have been theoretically predicted or experimentally confirmed to exhibit superconductivity with diverse coupling strengths. Notably, multiband superconductivity and anisotropic gap structures have emerged as defining characteristics of systems with complex Fermi surfaces, as exemplified by MgB$_2$~\cite{souma2003origin,liu2001beyond}

    The field gained significant momentum following the discovery of superconductivity in MgB$_2$ with a $T_c$ of 39 K \cite{nagamatsu2001superconductivity}, which established metal diborides (MB$_2$) as prototypical conventional superconductors. Subsequent advances in fabrication techniques further expanded this class, enabling the realization of two-dimensional MgB$_2$ with a reduced $T_c$ of approximately 31 K \cite{bekaert2017free}, as well as high-quality thick films exhibiting $T_c$ up to 36 K \cite{cheng2018fabrication}. These developments have revitalized interest in layered boride systems. Motivated by these findings, theoretical studies have predicted superconductivity in a variety of two-dimensional metal borides beyond MB$_2$. In particular, MB$_4$ compounds (M = Be, Mg, Ca, Sc, Al) have been proposed as promising candidates, with estimated $T_c$ values reaching 36.1 K for CaB$_4$, while AlB$_4$, BeB$_4$, and MgB$_4$ exhibit $T_c$ values of 30.9 K, 29.9 K, and 22.2 K, respectively \cite{sevik2022high} and its hydrogenated derivatives~\cite{seeyangnok2026stability}.

    Hydrogenation has emerged as an effective approach for tuning the electronic and superconducting properties of two-dimensional materials, particularly by enhancing electron--phonon coupling and increasing $T_c$. Following the experimental realization of Janus MoSH via the SEAR technique~\cite{lu2017janus}, this system was predicted to exhibit superconductivity with $T_c \approx 27$~K alongside possible charge-density-wave (CDW) instability~\cite{liu2022two,ku2023ab,jseeyang_cdw_moxh}, with further studies extending to its bilayer form~\cite{pinsook2025superconductivity}. This strategy has since been applied to a broader range of materials, including Ti$_2$CSH~\cite{jseeyang_ti2csh} and hydrogenated tungsten-based analogues with $T_c$ around 12~K~\cite{seeyangnok2024superconductivity,seeyangnok2024superconductivitywseh,qiao2024prediction,gan2024hydrogenation,fu2024superconductivity}. Similar superconducting behavior has also been predicted in group-IV transition-metal dichalcogenides (M = Ti, Zr, Hf; X = S, Se, Te) with $T_c$ in the range of 9–30~K~\cite{li2024machine}. However, many Janus MXH systems exhibit magnetic ground states~\cite{seeyangnok2025competition,sukserm2025half}, highlighting the complex interplay between magnetism and superconductivity in hydrogenated 2D materials.

    On the other hand, boron- and phosphorus-based compounds have attracted increasing attention due to their diverse bonding configurations and tunable electronic properties~\cite{oganov2009ionic,zhang2017two,carvalho2016phosphorene}. Boron is known to form electron-deficient networks with strong covalent bonding and multicenter interactions~\cite{oganov2009ionic,pancharatna2022anatomy}, whereas phosphorus introduces directional bonding and lone-pair effects that influence the electronic structure~\cite{liu2014phosphorene,xia2014rediscovering}. The combination of B and P atoms in a two-dimensional framework can lead to distinct bonding characteristics and enhanced orbital hybridization. These features may affect the electronic structure and electron--phonon interactions, providing a basis for further investigation of their physical properties.
    
    In this work, we investigate the structural, electronic, and superconducting properties of a hexagonal BP$_3$ monolayer using first-principles calculations and anisotropic Migdal--Eliashberg theory. Notably, this material has recently been explored in the context of sodium-ion battery applications~\cite{vu2026bp}, highlighting its potential multifunctionality. We show that the system is dynamically and thermally stable, exhibits intrinsic polarity, and possesses a multiband metallic electronic structure with strong $p$-orbital hybridization. The electron--phonon coupling is found to be substantial, leading to a superconducting state with a critical temperature of $T_c = 9.7$~K. Notably, the superconducting gap exhibits clear two-gap behavior, reflecting the multiband nature of the Fermi surface. Our results highlight BP$_3$ as a promising 2D superconductor and provide insight into the role of mixed bonding and multiband effects in enhancing superconductivity in low-dimensional materials.

\section{Computational Methods}

Density functional theory (DFT) calculations were carried out using the \textsc{Quantum ESPRESSO} (QE) package~\cite{giannozzi2009quantum}. A plane-wave basis set was employed with a kinetic energy cutoff of 80~Ry for the wavefunctions and 320~Ry for the charge density. Brillouin zone integrations were performed using a \(12 \times 12 \times 1\) Monkhorst--Pack \textit{k}-point mesh~\cite{monkhorst1976special}. Electronic occupations were treated using the Methfessel--Paxton smearing scheme with a smearing width of 0.02~Ry~\cite{methfessel1989high}. The electron--ion interaction was described by optimized norm-conserving Vanderbilt pseudopotentials~\cite{hamann2013optimized,schlipf2015optimization}, while exchange--correlation effects were treated within the generalized gradient approximation (GGA) using the Perdew--Burke--Ernzerhof (PBE) functional~\cite{perdew1996generalized}.

Structural relaxations were performed using the BFGS algorithm~\cite{liu1989limited} until the residual forces on each atom were less than \(10^{-5}~\text{eV/\AA}\). Phonon spectra and dynamical properties were calculated within density functional perturbation theory (DFPT) on a \(6 \times 6 \times 1\) \textit{q}-point grid. The electron--phonon interaction leads to finite phonon linewidths $\gamma_{\boldsymbol{q}\nu}$, and the corresponding mode-resolved electron--phonon coupling strength is given by
\begin{equation}
\lambda_{\boldsymbol{q}\nu} = 
\frac{\gamma_{\boldsymbol{q}\nu}}{\pi N(\epsilon_F)\,\omega_{\boldsymbol{q}\nu}^2},
\end{equation}
where \( N(\epsilon_F) \) is the electronic density of states at the Fermi level and \(\omega_{\boldsymbol{q}\nu}\) denotes the phonon frequency. 

Superconducting properties were investigated by solving the anisotropic Migdal--Eliashberg equations~\cite{migdal1958interaction,eliashberg1960interactions,nambu1960quasi,pinsook2024analytic} using the EPW code~\cite{noffsinger2010epw,ponce2016epw}, which relies on Wannier--Fourier interpolation~\cite{giustino2007electron,giustino2017electron}. The superconducting gap function \(\Delta_{nk}(i\omega_j)\) and the renormalization function \(Z_{nk}(i\omega_j)\) were determined self-consistently given by 
The Migdal--Eliashberg equations are given by
\begin{eqnarray}
Z_{nk}(i\omega_j) &=& 1 + \frac{\pi T}{N(\varepsilon_F)\omega_j} 
\sum_{mk'j'} 
\frac{\omega_{j'}}{\sqrt{\omega_{j'}^2 + \Delta_{mk'}^2(i\omega_{j'})}} \nonumber \\
&&\times \lambda(nk, mk', \omega_j - \omega_{j'})\,
\delta(\epsilon_{mk'} - \varepsilon_F),
\end{eqnarray}
\begin{eqnarray}
Z_{nk}(i\omega_j)\Delta_{nk}(i\omega_j) &=& 
\frac{\pi T}{N(\varepsilon_F)} 
\sum_{mk'j'} 
\frac{\Delta_{mk'}(i\omega_{j'})}{\sqrt{\omega_{j'}^2 + \Delta_{mk'}^2(i\omega_{j'})}}  \delta(\epsilon_{mk'} - \varepsilon_F) \nonumber  \\ &&
\times \left[\lambda(nk, mk', \omega_j - \omega_{j'}) - \mu^*\right],
\end{eqnarray}
on the Matsubara frequency axis, where \(\omega_j = (2j+1)\pi T\). The Coulomb pseudopotential was fixed at \(\mu^* = 0.1\).

To ensure convergence, dense \textit{k}- and \textit{q}-point meshes of \(120 \times 120 \times 1\) and \(60 \times 60 \times 1\), respectively, were used. A Fermi surface broadening of 0.40~eV and a Matsubara frequency cutoff of 1.00~eV were applied. The delta functions were approximated using Gaussian smearing, with widths of 0.40~eV for electrons and 0.5~meV for phonons.

\section{Results and Discussion}

\subsection{Crystal structure}
    \begin{figure}[ht]
        \centering
        \includegraphics[width=8.5cm]{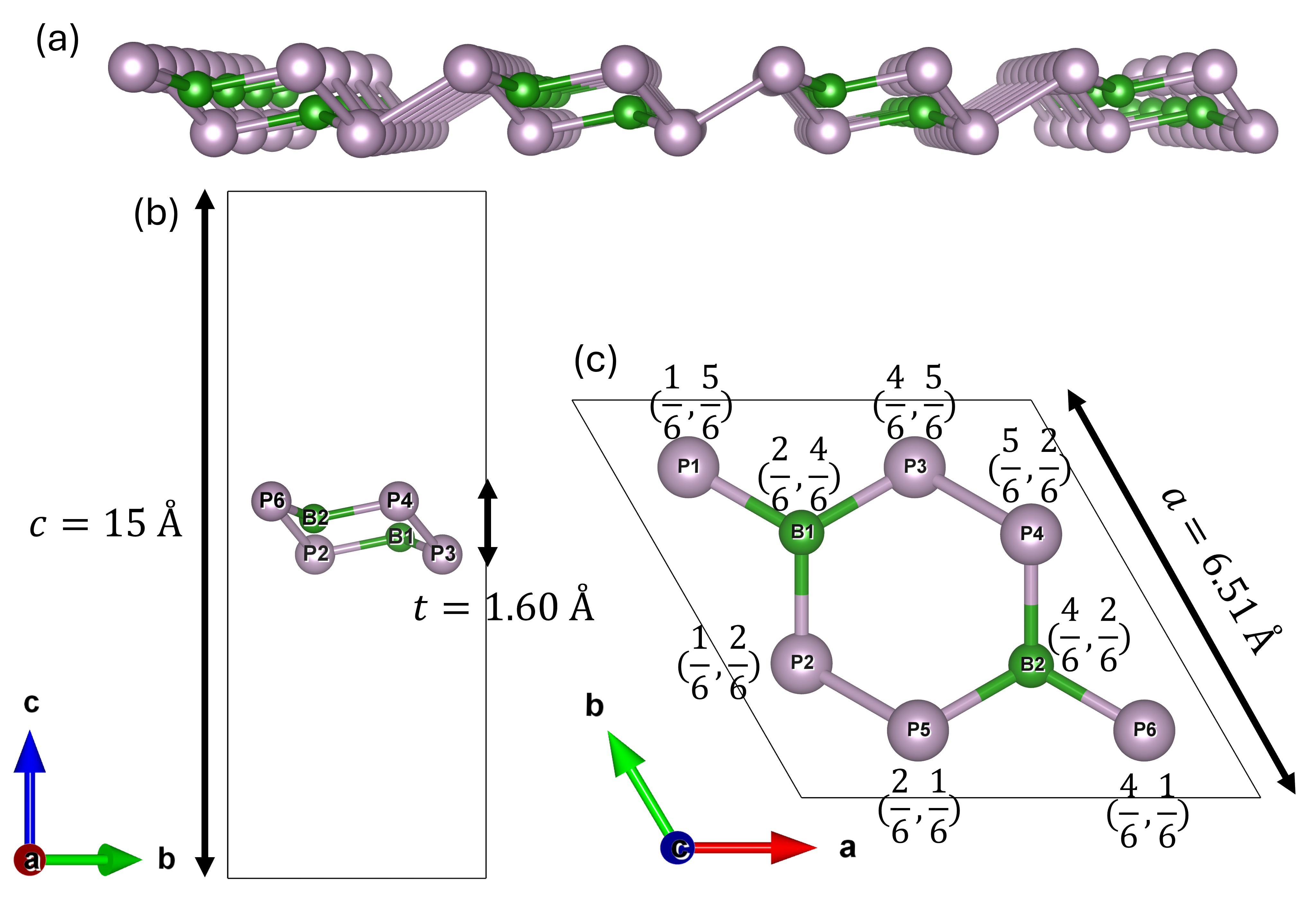}
        \caption{(a) Side view of the optimized monolayer structure, illustrating the buckled configuration of the P--B network. Purple and green spheres represent P and B atoms, respectively. (b) Simulation supercell consisting of an eight-atom monolayer used in this work, with a vacuum spacing of $c = 15$~\AA\ to avoid spurious interlayer interactions; the effective layer thickness is $t = 1.60$~\AA. (c) Top view of the primitive unit cell with lattice parameter $a = 6.51$~\AA. Atomic sites are labeled (P1--P6 and B1--B2), and their fractional coordinates are indicated.}
        \label{fig:structure}
    \end{figure}
    
    The optimized geometry of the hexagonal BP$_3$ monolayer is illustrated in Fig.~\ref{fig:structure}. As shown in Fig.~\ref{fig:structure}(a), the structure adopts a slightly buckled configuration rather than a perfectly planar arrangement, indicating the presence of mixed bonding characteristics within the lattice. The vertical displacement between atomic sublayers is quantified by an effective thickness of $t = 1.60$~\AA, which suggests a moderate out-of-plane. To eliminate artificial interlayer interactions, a vacuum spacing of $c = 15$~\AA\ is introduced along the out-of-plane direction, as depicted in Fig.~\ref{fig:structure}(b). The top view of the primitive unit cell in Fig.~\ref{fig:structure}(c) reveals a hexagonal-like framework composed of two inequivalent B atoms (B1 and B2) coordinated with six P atoms (P1–P6), forming a network of interconnected rings. The optimized lattice constant is found to be $a = 6.51$~\AA. The fractional atomic coordinates indicate a symmetric distribution of atoms within the unit cell, reflecting the underlying lattice symmetry. 
    
    \begin{figure}[h!]
        \centering
        \includegraphics[width=8.5cm]{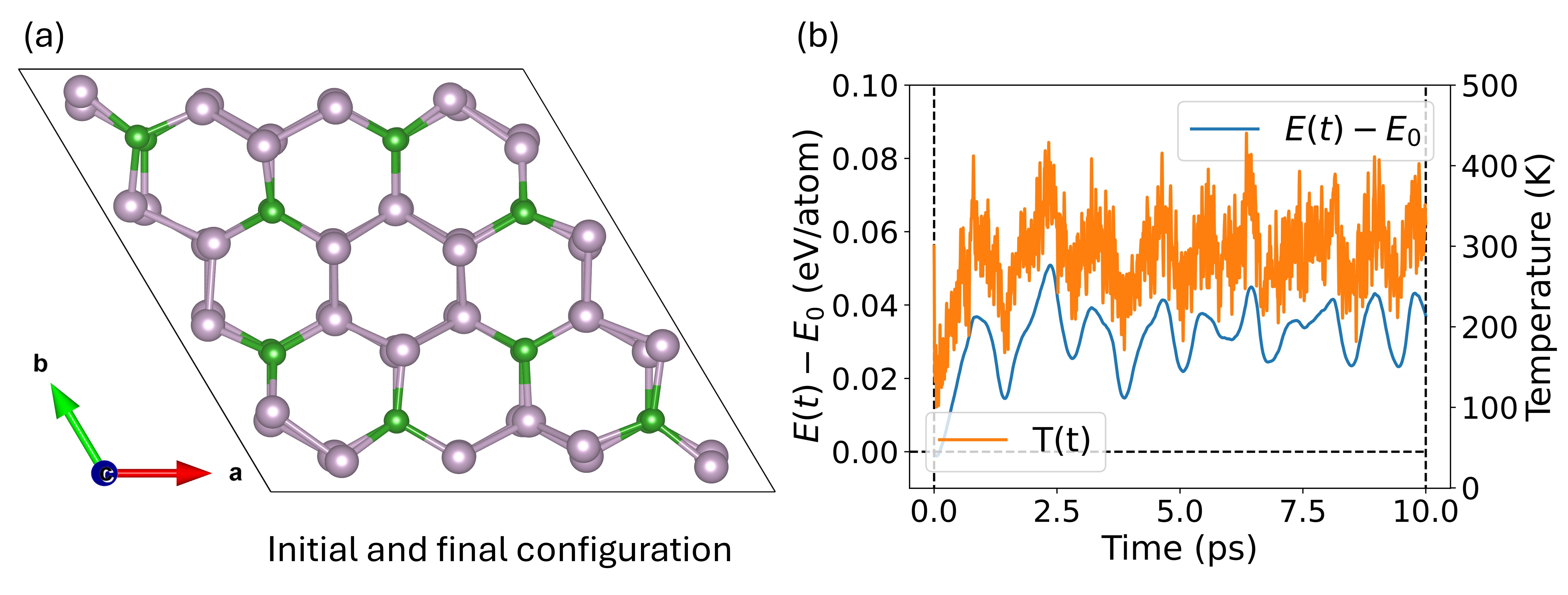}
        \caption{(a) Top view of the monolayer structures, showing the initial (lower structure) and final atomic configurations (upper strucutre) obtained from \textit{ab initio} molecular dynamics (AIMD) simulations. The structure maintains its integrity without noticeable reconstruction, indicating good thermal stability. Purple and green spheres denote P and B atoms, respectively. (b) Time evolution of the total energy difference per atom, $E(t)-E_0$, (left axis) and temperature $T(t)$ (right axis) during the AIMD simulation. The system exhibits small energy fluctuations and stable temperature around the target value, confirming the dynamical stability of the monolayer at finite temperature over the simulation time of 10~ps.}
        \label{fig:aimd}
    \end{figure}
    To assess the thermal stability of the proposed monolayer, \textit{ab initio} molecular dynamics (AIMD) simulations were performed, as summarized in Fig.~\ref{fig:aimd}. As shown in Fig.~\ref{fig:aimd}(a), the atomic structure remains intact throughout the simulation, with no evidence of bond breaking or structural reconstruction, indicating robust stability at finite temperature. The evolution of the total energy and temperature as a function of simulation time is presented in Fig.~\ref{fig:aimd}(b). After an initial equilibration period, the system reaches a steady state characterized by small fluctuations in both energy and temperature. The total energy oscillates around a constant average value, while the temperature remains close to the target value with only moderate thermal fluctuations, demonstrating effective thermal equilibration. These results confirm that the monolayer is dynamically stable and can maintain its structural integrity under thermal conditions. The mechanical stability of the hexagonal BP$_3$ monolayer, as previously reported~\cite{vu2026bp}, with $C_{11} = 115.98$~N/m, $C_{12} = 21.46$~N/m, and $C_{66} = 47.26$~N/m, satisfies the 2D mechanical stability criteria, $C_{11}C_{22} - C_{12}^2 > 0$ and $C_{11}, C_{22}, C_{66} > 0$~\cite{mouhat2014necessary}. These results confirm that the BP$_3$ monolayer is mechanically stable and potentially synthesizable.

\subsection{Electronic properties}
    \begin{figure}[h!]
        \centering
        \includegraphics[width=8.5cm]{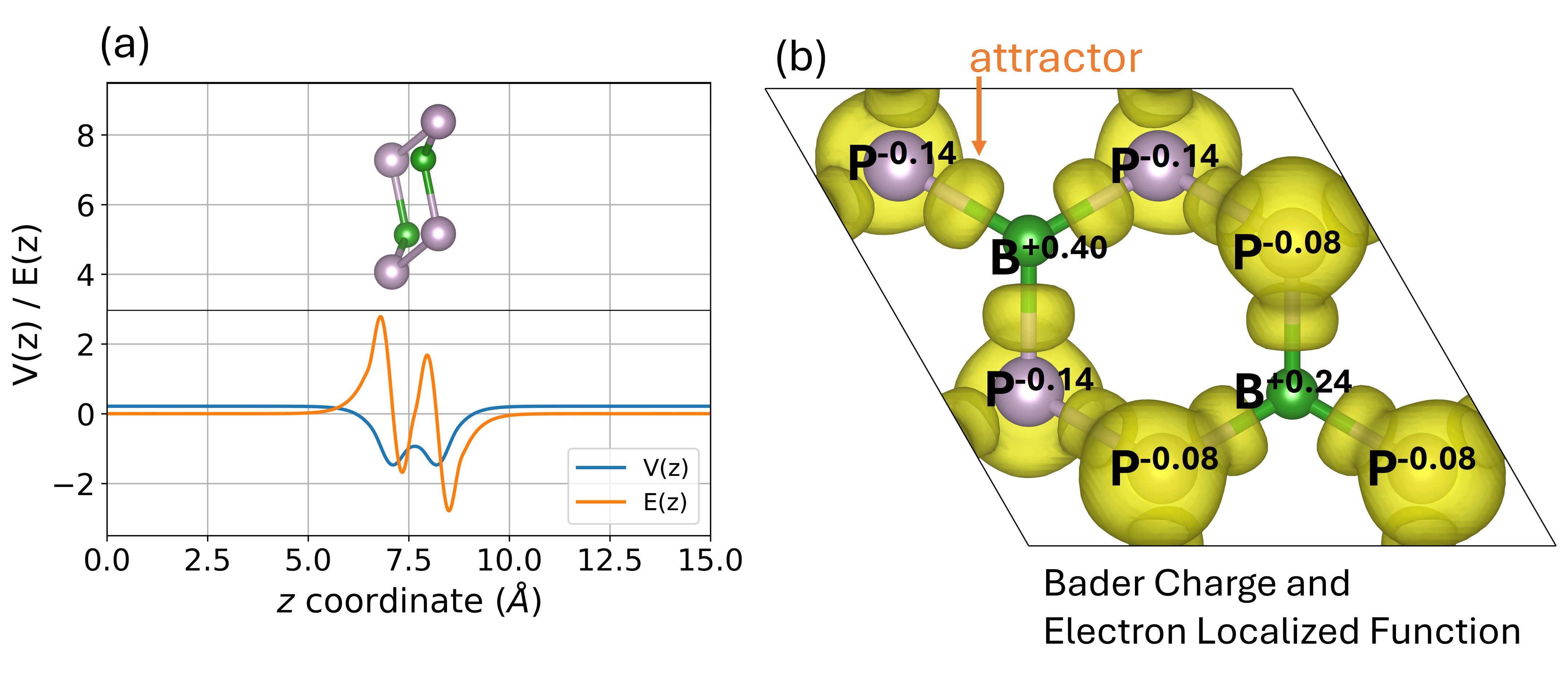}
    \caption{(a) Plane-averaged electrostatic potential $V(z)$ and the corresponding electric field $E(z)$ along the out-of-plane ($z$) direction of the monolayer. The asymmetric profile of $V(z)$ and the presence of a nonzero internal electric field suggest a built-in polarity across the layer. (b) Bader charge analysis combined with the electron localization function (ELF) isosurface. A small charge transfer from B to P atoms is observed, with B atoms carrying positive charges and P atoms acquiring negative charges. The presence of ELF attractors along the B–P bonds indicates covalent bonding between B and P atoms. The ELF distribution (yellow isosurface) highlights electron localization around P atoms, while reduced localization near B atoms suggests electron depletion.}
    \label{fig:elf_bader}
    \end{figure}

    The electrostatic properties and charge distribution of the monolayer are analyzed in Fig.~\ref{fig:elf_bader}. The plane-averaged electrostatic potential \(V(z)\) and the corresponding electric field \(E(z)\), shown in Fig.~\ref{fig:elf_bader}(a), exhibit a pronounced asymmetry along the out-of-plane direction. This asymmetric profile gives rise to a nonvanishing internal electric field, indicating the presence of an intrinsic dipole moment across the layer. Such built-in polarization originates from the structural asymmetry.

    Further insight into the bonding characteristics is provided by the Bader charge analysis and electron localization function (ELF), as shown in Fig.~\ref{fig:elf_bader}(b). A net charge transfer from B to P atoms is observed, with B atoms carrying positive charges (up to \(+0.40\,e\)) and P atoms acquiring negative charges (down to \(-0.14\,e\)), indicating partial ionic character. More importantly, the ELF distribution reveals pronounced electron localization regions (attractors) not only around the P atoms but also along the B--P bonds. These attractors signify the presence of covalent bonding with a significant \(\pi\)-bonding component arising from orbital overlap perpendicular to the plane. The accumulation of electron density in these regions highlights the coexistence of ionic charge transfer and directional covalent \(\pi\)-bonding within the lattice. Such mixed bonding character is crucial for understanding the electronic structure and may play a significant role in governing the transport and electron--phonon interaction properties of the monolayer.

    \begin{figure}[ht]
        \centering
        \includegraphics[width=8.5cm]{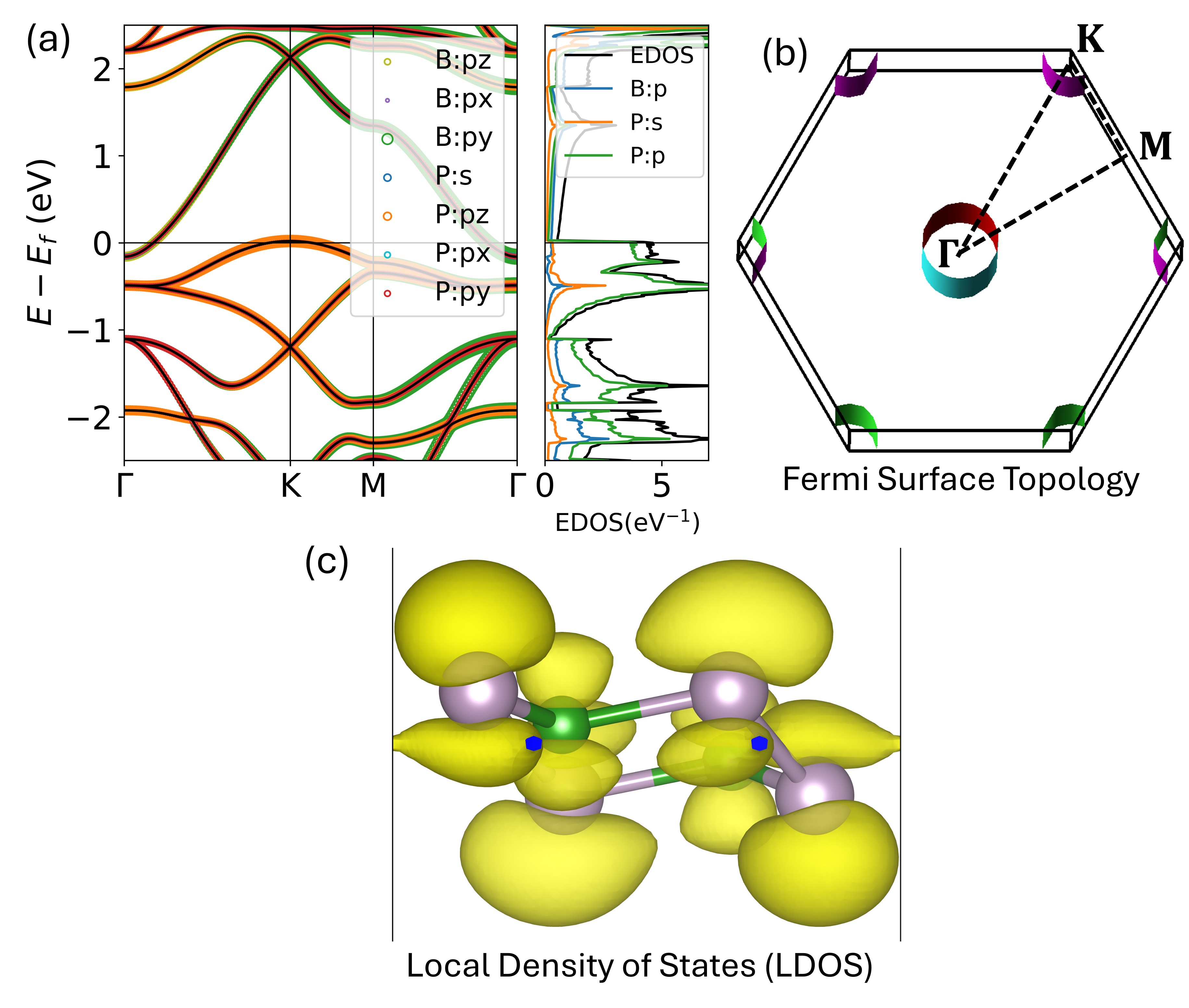}
        \caption{(a) Electronic band structure along the high-symmetry path $\Gamma$--K--M--$\Gamma$ together with the projected density of states (PDOS). The orbital contributions from B ($p_x$, $p_y$, $p_z$) and P ($s$, $p_x$, $p_y$, $p_z$) states are shown, indicating that the states near the Fermi level are predominantly derived from $p$ orbitals, particularly $p_z$, with strong hybridization between B and P atoms. (b) Fermi surface topology showing two distinct sheets corresponding to contributions from B ($p_z$) and P ($p_z$) orbitals, associated with inner and outer Fermi surfaces. (c) Local density of states (LDOS) distribution near the Fermi level, illustrating that the electronic states at the Fermi surface are dominated by $p_z$ orbitals, consistent with the projected band structure.}
    \label{fig:band_fermi_ldos}
    \end{figure}

    The electronic structure of the monolayer is presented in Fig.~\ref{fig:band_fermi_ldos}. As shown in Fig.~\ref{fig:band_fermi_ldos}(a), several bands cross the Fermi level, confirming the metallic nature of the system. The projected density of states (PDOS) indicates that the electronic states near the Fermi level are predominantly derived from $p$ orbitals of both B and P atoms, with a dominant contribution from $p_z$ orbitals and noticeable hybridization between them. The coexistence of in-plane ($p_x$, $p_y$) and out-of-plane ($p_z$) components suggests a mixed bonding character with a significant $\pi$-orbital contribution.

    The corresponding Fermi surface, shown in Fig.~\ref{fig:band_fermi_ldos}(b), consists of two distinct sheets associated with contributions from B ($p_z$) and P ($p_z$) orbitals, corresponding to inner and outer Fermi surfaces. This clearly reflects the multiband nature of the system. The anisotropic shape of these sheets suggests direction-dependent electronic transport and scattering processes, which may influence the superconducting behavior.
    
    Further insight is obtained from the local density of states (LDOS) depicted in Fig.~\ref{fig:band_fermi_ldos}(c). The LDOS isosurfaces show electron accumulation along the B--P bonds and distinct lobes extending perpendicular to the plane, characteristic of $p_z$-dominated states. This spatial distribution indicates that the electronic states near the Fermi level are primarily derived from $p_z$ orbitals, consistent with the projected band structure. Overall, the multiband electronic structure and orbital hybridization provide a suitable framework for electron--phonon interactions in this system.

\subsection{Phonons and Electron-phonon couplings}
\begin{figure}[ht]
    \centering
    \includegraphics[width=8.5cm]{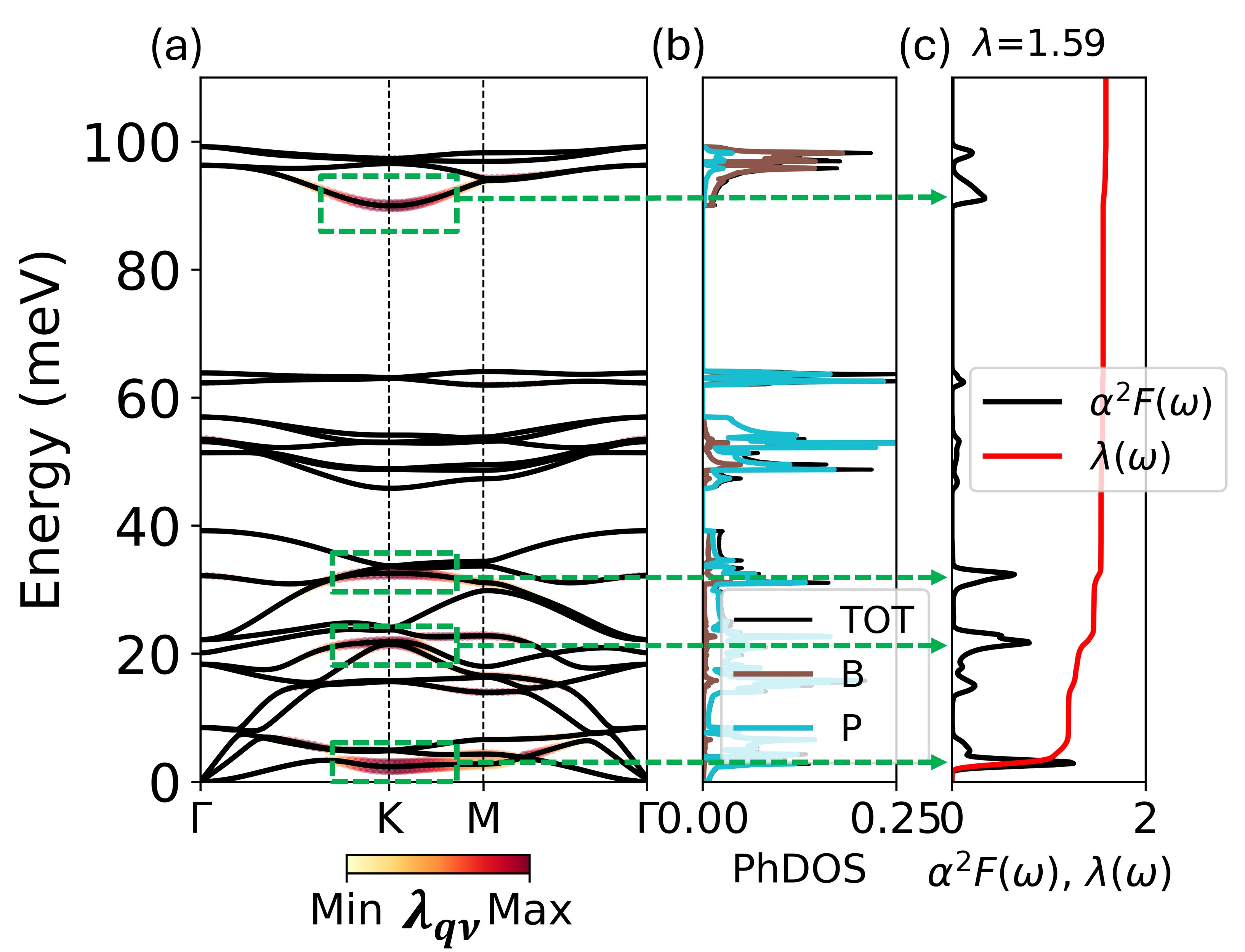}
    \caption{(a) Electron--phonon coupling (EPC)-weighted phonon dispersion along the high-symmetry path $\Gamma$--K--M--$\Gamma$, where the color scale represents the magnitude of the EPC strength; highlighted regions indicate phonon modes with significant coupling contributions. (b) Phonon density of states (PhDOS), decomposed into contributions from B and P atoms. (c) Eliashberg spectral function $\alpha^2F(\omega)$ (black curve) and the cumulative electron--phonon coupling constant $\lambda(\omega)$ (red curve). The total EPC constant is $\lambda = 1.59$, indicating strong coupling behavior.}
    \label{fig:epc}
    \end{figure}

    The electron--phonon coupling (EPC) characteristics of the monolayer are summarized in Fig.~\ref{fig:epc}. The EPC-weighted phonon dispersion shown in Fig.~\ref{fig:epc}(a) reveals that the coupling strength is not uniformly distributed across phonon modes, but is instead concentrated in specific branches near the K and M points, particularly in the ranges of 0--40~meV and 90--100~meV. These modes, highlighted by higher color intensity, indicate the dominant scattering channels contributing to the EPC. Notably, soft acoustic and low-energy optical phonons near the high-symmetry K and M points play a crucial role in enhancing the overall coupling strength.

    The phonon density of states (PhDOS), presented in Fig.~\ref{fig:epc}(b), provides further insight into the atomic origin of these vibrational modes. The spectrum shows significant contributions from both B and P atoms over a broad energy range. The low-frequency modes are primarily associated with the heavier atomic motions of phosphorus, while the high-frequency modes arise from the lighter boron atoms, involving both in-plane and out-of-plane vibrations and reflecting the stiffer bonding due to their covalent character.

    The Eliashberg spectral function $\alpha^2F(\omega)$ and the accumulated EPC constant $\lambda(\omega)$, shown in Fig.~\ref{fig:epc}(c), clearly demonstrate that the main contribution to the total EPC constant ($\lambda = 1.59$) originates from both low- and high-frequency phonon modes. The pronounced peaks in $\alpha^2F(\omega)$ coincide with the strongly coupled phonon branches identified in the dispersion, confirming their dominant role in mediating superconductivity. The low-frequency modes contribute up to 94\% of the total EPC, whereas the high-frequency modes in the range of 90--100~meV account for only about 6\% of the total coupling. The rapid increase of $\lambda(\omega)$ within these frequency ranges indicates that these modes provide the largest contribution to the pairing interaction.

\subsection{Anisotropic Superconductivity}
\begin{figure}[ht]
    \centering
    \includegraphics[width=8.5cm]{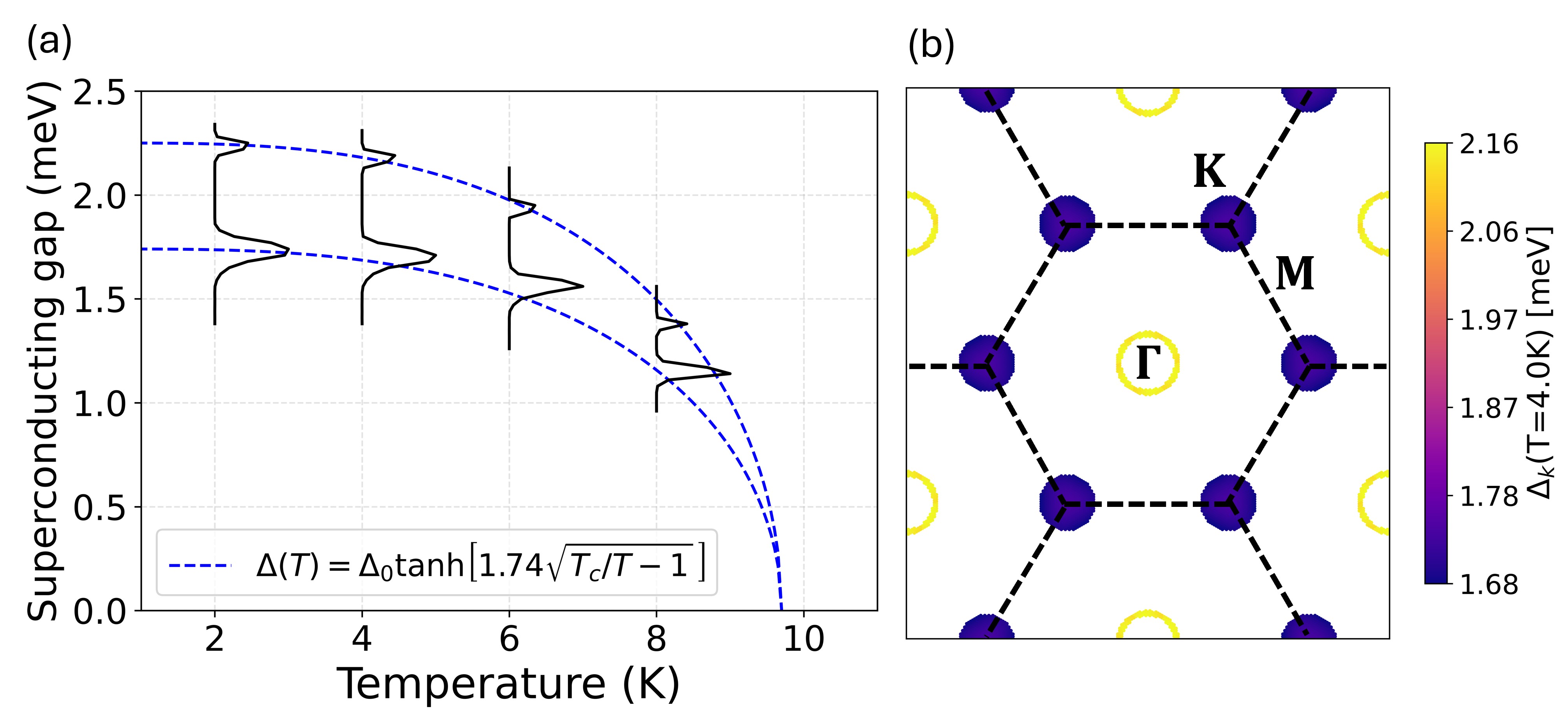}
    \caption{(a) Temperature dependence of the superconducting gap $\Delta(T)$ obtained from anisotropic Migdal--Eliashberg calculations. The dashed blue line represents the BCS fit, $\Delta(T) = \Delta_0 \tanh!\left[1.74\sqrt{T_c/T - 1}\right]$, yielding gap values of $\Delta_0 = 2.25$ and $1.74$~meV at 2~K, with a critical temperature of $T_c = 9.7$~K. The black curves illustrate the distribution of gap values on the Fermi surface at 2~K, indicating moderate anisotropy and clear two-gap superconductivity. (b) Momentum-resolved superconducting gap $\Delta_{\mathbf{k}}$ on the Fermi surface at $T = 4.0$~K, showing that the two-gap superconductivity originates from two distinct Fermi surface sheets associated with the $p_z$ orbitals of B and P.}
    \label{fig:gap}
    \end{figure}

    The superconducting gap structure shown in Fig.~\ref{fig:gap} reveals clear signatures of multiband (two-gap) superconductivity. As illustrated in Fig.~\ref{fig:gap}(a), the gap distribution exhibits multiple distinct values at a given temperature, rather than collapsing into a single uniform gap. This behavior originates from the presence of two Fermi surface sheets with different orbital characters and electron--phonon coupling strengths. In particular, bands with stronger coupling give rise to a larger gap associated with the $p_z$ orbitals of boron, while more weakly coupled bands sustain a smaller gap associated with the $p_z$ orbitals of phosphorus, leading to a two-gap scenario. At low temperature ($T = 2$~K), two distinct gap values of approximately 2.25 and 1.74~meV are observed, consistent with the multiband nature of the electronic structure and the anisotropic distribution of EPC strength.

    The temperature dependence of the superconducting gap follows the general trend predicted by BCS theory, as indicated by the agreement with the fitted curve. Using the larger gap value $\Delta_0^{(L)} \approx 2.25$~meV and $T_c = 9.7$~K yields a gap ratio $2\Delta_0^{(L)} / k_B T_c \approx 5.6$, while the smaller gap $\Delta_0^{(S)} \approx 1.74$~meV gives $2\Delta_0^{(S)} / k_B T_c \approx 4.3$. Both values exceed the weak-coupling BCS limit of 3.53, indicating strong-coupling superconductivity. The smooth closing of the gap near $T_c$ is consistent with a phonon-mediated pairing mechanism.

    The momentum-resolved gap distribution in Fig.~\ref{fig:gap}(b) provides further insight into the anisotropic superconducting state. The gap remains finite across the entire Fermi surface, confirming a fully gapped (nodeless) state, while its variation reflects differences between distinct Fermi surface sheets. The two-gap behavior originates from two separate sheets associated with the $p_z$ orbitals of B and P atoms. Overall, the system can be described as a strongly coupled, anisotropic two-gap superconductor with a moderately high critical temperature for a two-dimensional material. These results place the hexagonal BP$_3$ monolayer among known two-gap superconductors—such as n-doped graphene~\cite{margine2014two}, AlB$_2$-based thin films~\cite{zhao2019two}, trilayer LiB$_2$C$_2$~\cite{gao2020strong}, monolayer LiBC~\cite{modak2021prediction}, GaInSLi~\cite{seeyang2025phase_japtwpgap}, MoSLi~\cite{xie2024strong}, MoSeLi~\cite{seeyangnok2026tunable}, MoSH~\cite{liu2022two}, and hydrogenated borides (MgB$_4$H, CaB$_4$H, and AlB$_4$H)~\cite{seeyangnok2026stability}, highlighting its intrinsic multigap superconducting nature.

\section*{Conclusions}
    In this work, we have systematically investigated the structural, electronic, and superconducting properties of the hexagonal BP$_3$ monolayer using first-principles calculations and anisotropic Migdal--Eliashberg theory. The optimized structure exhibits a slightly buckled geometry and is dynamically and thermally stable, as confirmed by phonon calculations and \textit{ab initio} molecular dynamics simulations, indicating its feasibility for experimental realization. The bonding characteristics, including orbital hybridization and $\pi$-type contributions, play an important role in determining the electronic properties.

    The electronic structure shows metallic behavior with multiple bands crossing the Fermi level, leading to a multiband Fermi surface with significant contributions from $p_z$ orbitals of B and P atoms. This multiband nature is reflected in the superconducting properties. The phonon spectrum confirms dynamical stability, and the electron--phonon coupling analysis yields a relatively large coupling constant $\lambda = 1.59$, with dominant contributions from low- and intermediate-frequency phonon modes.

    Solving the anisotropic Migdal--Eliashberg equations yields a superconducting critical temperature of $T_c = 9.7$~K. The superconducting state is nodeless but anisotropic, with two distinct gap values of approximately 2.25 and 1.74~meV at low temperature, confirming two-gap superconductivity arising from distinct Fermi surface sheets associated with B and P $p_z$ orbitals. Overall, the BP$_3$ monolayer can be described as a strongly coupled, anisotropic two-gap superconductor with a moderately high critical temperature for a two-dimensional system. These results provide insight into superconductivity in boron–phosphorus-based materials and motivate further theoretical and experimental investigations of related low-dimensional systems.

\section*{Acknowledgments}
	This research project is supported by the Second Century Fund (C2F), Chulalongkorn University. We acknowledge the supporting computing infrastructure provided by NSTDA, CU, CUAASC, NSRF via PMUB [B05F650021, B37G660013] (Thailand). (\url{URL:www.e-science.in.th}).

\bibliography{references}

\end{document}